\documentclass[12pt]{article}
\usepackage[english]{babel}
\usepackage[utf8]{inputenc}
\usepackage[T1]{fontenc}

\def\mF{\mathcal{F}}
\usepackage{amsmath}
\def\tPi{\tilde{\Pi}}

\usepackage{amsfonts}
\usepackage{url}
\usepackage[dvips]{graphicx}
\usepackage{calc}
\usepackage{subfigure}
\usepackage{multirow}

\def\mL{\mathcal{L}}

\def\mH{\mathcal{H}}

\def\bx{\mathbf{x}}
\newcommand{\pb}[1]{\left\{#1\right\}}

\def\mG{\mathcal{G}}
\def\by{\mathbf{y}}

\def\bT{\mathbf{T}}

\begin{document}
	\begin{center}
		{\Large{ \bf Canonial Analysis of General Relativity Formulated with the 
		New Metric $f^{ab}=(-g)^{\alpha}g^{ab}$}}
		
		\vspace{1em}  J. Kluso\v{n}
		\\
		Email address:
		 klu@physics.muni.cz\\
		
		\vspace{1em}  \textit{Department of Theoretical Physics and
			Astrophysics, Faculty
			of Science,\\
			Masaryk University, Kotl\'a\v{r}sk\'a 2, 611 37, Brno, Czech Republic}\\		
	\end{center}
	
	\abstract{In this short note we investigate canonical formalism for General Relativity which is formulated with the metric $f^{ab}=(-g)^\alpha g^{ab}$.
	We find corresponding Hamiltonian and we show that constraint structure is
	the same as in the standard formulation.}	
		
			\newpage

\section{Introduction and Summary}\label{first}
Dynamical variable of General Relativity is metric with components  $g_{ab}$. They are natural variables for formulation of Riemann geometry and corresponding quantities as for example scalar curvature that is fundamental part of Einstein-Hilbert action. Properties of this action were carefully examined by T. Padmanabhan in 
\cite{Parattu:2013gwa,Padmanabhan:2013nxa}. However it was stressed here that it is possible to define  new variables $f^{ab}=\sqrt{-g}g^{ab}$ which could be even more appropriate  for description of dynamics of gravity.
 An importance of these variables was already stressed in \cite{Edington,Schrodinger,Einstein:1955ez} and were recently presented in an important paper  \cite{Padmanabhan:2013nxa}. In fact, they have significant meaning in covariant canonical formulation \cite{DeDonder,Weyl}
 of General Relativity 
 \cite{Horava:1990ba}
 \footnote{For review, see for example \cite{Struckmeier:2008zz,Kastrup:1982qq}.}
  and its generalization 
  \cite{Kanatchikov:2023yde,Kluson:2023idq,Kluson:2022qxl,Kluson:2020tzn,Riahi:2019tyb,Rovelli:2002ef,Kanatchikov:1993rp}.  
 Further, it was nicely shown in \cite{Parattu:2013gwa} that these variables have nice thermodynamic interpretation in the emergent gravity paradigm that claims that gravity is emergent from some unknown more fundamental theory. Then we could conjecture that $f^{ab}$ variables are more fundamental than $g_{ab}$ and study consequence of this hypothesis. In particular, it would be interesting to formulate canonical formalism 
 for these variables. By canonical formalism we mean conventional formalism based on $D+1$ splitting of space time
\cite{Arnowitt:1962hi}, for review see \cite{Gourgoulhon:2007ue}. The goal of this paper is investigate this question.

More precisely, we consider theory with the metric $f^{ab}$ that is related to $g^{ab}$ by point transformation $f^{ab}=(-g)^\alpha g^{ab}$, where parameter $\alpha$ can be arbitrary number  and our goal is to study dependence of the theory on $\alpha$. Then in order to find canonical formulation of theory for $f^{ab}$ variable we should again perform $D+1$ splitting of $f^{ab}$ when we introduce variables $M,a^{ij},M^i$ whose precise definitions will be given in the next section and   which are related to similar splitting of $g^{ab}$ metric in terms of $N,h_{ij},N^i$. With the help of these relations we will be able to find corresponding conjugate momenta. During this procedure we also find primary constraint that relates $a^{ij}$ with $M$ and which is a consequence of the fact that $M$ is dynamical variable as opposite to the lapse $N$ whose conjugate momentum is primary constraint of the theory in the original formulation. On the other hand performing standard manipulation we obtain Hamiltonian that has similar form as the standard one. Then the requirement of the preservation of the primary constraints leads to emergence of $D+1$ secondary constraints which are Hamiltonian constraint together with $D$ spatial diffeomorphism constraints. Finally we study stability of these constraints. It turns out that it is useful to express them in terms of the original variables as composite objects from $a^{ij},M$ and conjugate momenta. Then it turns out that the constraints and Poisson brackets between them have the same form as in General Relativity. 

We also consider the case when we use new set of variables for spatial metric $h_{ij}$ only. In this case the situation is simpler than in the more general case since lapse function does not change. We determine corresponding Hamiltonian and  constraint structure that has again the same form as in General Relativity. Finally we argue that this Hamiltonian can be derived by gauge fixing of the primary constraint in the model with dynamical $M$.  

Let us outline our results. We investigate General Theory formulated in terms of new variable $f^{ab}$ and study their constraint structure. We show that compared to the original case the Hamiltonian is more complicated and introducing new variables does not bring new benefits for the theory. In other words while $f^{ab}$ variable has significant meaning in the thermodynamics interpretation of the theory and in the covariant canonical formalism standard Hamiltonian formalism is naturally formulated in terms of original metric $g_{ab}$ and conjugate momenta. 

This paper is organized as follows. In the next section (\ref{second}) we introduce new variable $f^{ab}$ and determine corresponding Hamiltonian. In the section (\ref{third}) we study stability of the primary constraints and determine constraint structure of theory. Finally in section (\ref{fourth}) we introduce new variable for spatial part of the metric only and study corresponding Hamiltonian. 

\section{Hamiltonian Formalism for $f^{ab}$ metric}\label{second}
As we wrote in the introduction the  goal of this paper is to study canonical structure of General Relativity action 
which is expressed in terms  of variable  $f^{ab}=(-g)^\alpha g^{ab}$ in $D+1$ dimensions, where $a,b=0,1,\dots,D$. This form of the metric is generalization of the relation 
$f^{ab}=\sqrt{-g}g^{ab}$ that was introduced long time ago by Einstein and whose 
importance in the covariant canonical formalism was stressed recently by Padmanabhan in 
\cite{Parattu:2013gwa}. An unanswered question remains what is the form of the canonical structure for the General Relativity formulated with the variable $f^{ab}$. The goal of this paper is to find Hamiltonian for  $f^{ab}$ and conjugate momenta. 

In the first step of our analysis we  find inverse relations between $g^{ab}$ and $f^{ab}$. From definition above we get
\begin{equation}
	-f=(-g)^{\alpha(D+1)}(-g)^{-1} \ , f\equiv \det f^{ab} \ , g\equiv \det g_{ab} \ . 
\end{equation}
With the help of these results we find inverse relation 
\begin{equation}
	(-g)=(-f)^{\frac{1}{\alpha (D+1)-1}} \ , \quad  g^{ab}=f^{ab}(-f)^{-\frac{\alpha}{\alpha(D+1)-1}} \ . 
\end{equation}
In order to find canonical formulation of theory we  again presume $D+1$ decomposition 
of metric $g_{ab}$. Explicitly, in case of metric $g_{ab}$ we introduce 
lapse function $N=1/\sqrt{-g^{00}}$ and the shift function
$N^i=-g^{0i}/g^{00}$. In terms of these variables we
write the components of the metric $g_{ab}$ as
\begin{eqnarray}
	g_{00}=-N^2+N_i h^{ij}N_j \ , \quad g_{0i}=N_i \ , \quad
	g_{ij}=h_{ij} \ ,
	\nonumber \\
	g^{00}=-\frac{1}{N^2} \ , \quad g^{0i}=\frac{N^i}{N^2} \
	, \quad g^{ij}=h^{ij}-\frac{N^i N^j}{N^2} \ ,
	\nonumber \\
\end{eqnarray}
where $h_{ij}$ is non-singular spatial $D-$dimensional metric with inverse
$h^{jk}$ so that $h_{ij}h^{jk}=\delta_i^k$.

Let us presume the same decomposition of $f^{ab}$
and its inverse $f_{ab}$
\begin{eqnarray}
	f_{00}=-M^2+M_i a^{ij}M_j \ , \quad f_{0i}=M_i \ , \quad 
	f_{ij}=a_{ij} \ , \nonumber \\
	f^{00}=-\frac{1}{M^2} \ , \quad f^{0i}=\frac{M^i}{M^2} \ , 
	\quad f^{ij}=a^{ij}-\frac{M^iM^j}{M^2} \ , 
	\nonumber \\
\end{eqnarray}
where $a_{ij}a^{jk}=\delta_i^k$ and where $M^i=a^{ij}M_j$. 
Then if we firstly compare $g^{00}$ with $f^{00}$ we obtain
\begin{eqnarray}
	\frac{1}{N^2}=\frac{1}{M^2}(M^2a)^{\frac{\alpha}{\alpha(D+1)-1}} \ ,  \quad 
	a\equiv \det a_{ij} \ . 
	\nonumber \\
\end{eqnarray}
As the next step we consider relation between $g^{0i}$ and $f^{0i}$ and we get
\begin{eqnarray}
	N^i=M^i\frac{N^2}{M^2}(M^2a)^{\frac{\alpha}{\alpha(D+1)-1}}=M^i \ , \nonumber \\
\end{eqnarray}
that implies equality of the shift functions 
\begin{equation}
N^i=M^i \ . 
\end{equation}	
Finally we proceed to the relation between $g^{ij}$ and $f^{ij}$ that allows us to find relation between $h^{ij}$ and $a^{ij}$ in the form 	
\begin{eqnarray}	
h^{ij}=
a^{ij}(M^2a)^{\frac{\alpha}{\alpha(D+1)-1}} \ .
	\nonumber \\
\end{eqnarray}
In summary, we have following relations between original and new variables
\begin{eqnarray}
	\frac{1}{N^2}=\frac{1}{M^2}(M^2a)^{\frac{\alpha}{\alpha(D+1)-1}}  \ , \quad 
	N^i=M^i \ , \quad 
	h^{ij}=a^{ij}(M^2a)^{\frac{\alpha}{\alpha(D+1)-1}} \ . \nonumber \\
\end{eqnarray}
Now we are ready to proceed to the definition of the momenta conjugate to $M,M^i$ and $a_{ij}$. 
Note that the action for gravity in $D+1$ dimensions has the form 
\begin{equation}\label{Sr}
S=\frac{1}{\kappa}\int d^{D+1}x\sqrt{-g}R \ ,  \quad \kappa=16\pi G \ , 
\end{equation}
where $R(g)$ is scalar curvature. In  order to find canonical action 
we use $D+1$ decomposition of $R$ 
\begin{eqnarray}\label{Rexpan}
&&R=K_{ij}\mG^{ijkl}K_{kl}+r(h)+\frac{2}{\sqrt{-g}}\partial_\mu[\sqrt{-g}n^\mu K]
-\frac{2}{\sqrt{h}N}\partial_i[\sqrt{h}h^{ij}\partial_j N] \ , 
\nonumber \\
&&K_{ij}=\frac{1}{2N}(\partial_t h_{ij}-\nabla_i N_j-\nabla_j N_i) \ , \nonumber \\
&& n^0=\sqrt{-g^{00}} \ , \quad 
n^i=-\frac{g^{0i}}{\sqrt{-g^{00}}} \ , \nonumber \\
\end{eqnarray}
and where $r$ is scalar curvature defined with $h_{ij}$ and $\nabla_i$ is covariant
derivative compatible with the metric $h_{ij}$. Note that the divergence terms in (\ref{Rexpan}) can be ignored in the action (\ref{Sr}).
Finally we introduced  de Witt metric $\mG^{ijkl}$ and its inverse $\mG_{ijkl}$ that have the form 
\begin{eqnarray}
&&	\mG^{ijkl}=\frac{1}{2}(h^{ik}h^{jl}+h^{il}h^{jk})-h^{ij}h^{kl}  \ , \nonumber \\
&&	\mG_{ijkl}=\frac{1}{2}(h_{ik}h_{jl}+h_{il}h_{jk})-\frac{1}{D-2}h_{ij}h_{kl} \ . \nonumber \\
\end{eqnarray}	
From their definition we obtain useful relations	
\begin{equation}	
h_{ij}\mG^{ijkl}=-(D-2)h^{kl} \ , \quad 
h^{ij}\mG_{ijkl}=-\frac{1}{D-2}h_{kl} \ . 
\end{equation}
As the first step we proceed with the momentum conjugate to $M$ and we get
\begin{eqnarray}
&&	p_M=\frac{\partial \mL}{\partial (\partial_0 M)}
	=\frac{\partial \mL}{\partial (\partial_0N)}\frac{\partial (\partial_0 N)}{
		\partial(\partial_0 M)}+
	\frac{\partial \mL}{\partial (\partial_0 h_{ij})}\frac{\partial (\partial_0 h_{ij})}{
		\partial(\partial_0 M)}=\nonumber \\
&&	=\frac{\partial \mL}{\partial (\partial_0N)}\frac{\partial  N}{
		\partial M}+
	\frac{\partial \mL}{\partial (\partial_0 h_{ij})}\frac{\partial  h_{ij}}{
		\partial M}=\nonumber \\
&&	=\frac{2\alpha(D-2)}{\kappa M(\alpha(D+1)-1)}\sqrt{h}K \ , \nonumber \\
	\nonumber \\
\end{eqnarray}
where we used  the fact that $f^{ab}$ and $g^{ab}$ are related by point transformations so that 
\begin{equation}
	\frac{\partial (\partial_0 g^{ab})}{\partial (\partial_0 f^{cd})}=
	\frac{\partial g^{ab}}{\partial f^{cd}}
\end{equation}
and also the fact $h^{ij}h_{jk}=\delta^i_k$ so that
\begin{equation}
	\frac{\partial h_{ij}}{\partial M}=-h_{im}\frac{\delta h^{mn}}{\delta M}
	h_{nj} \ . 
\end{equation}
In case of the momenta conjugate to $M^i$ we also get that they are primary constraints 
\begin{eqnarray}
	\pi_i=\frac{\partial \mL}{\partial (\partial_0M^i)}\approx 0 
\end{eqnarray}
as follows from the fact that  $M^i=N^i$ and from the fact that the action (\ref{Sr}) does not depend on time derivative of $N^i$. 

 Finally we proceed to the momentum conjugate to $a^{ij}$ and we get
\begin{eqnarray}
&&	\pi_{ij}=\frac{\partial \mL}{\partial (\partial_0 a_{ij})}=
	\frac{\partial \mL}{\partial (\partial_0 N)}\frac{\partial (\partial_0 N)}{\partial(\partial_0 a^{ij})}+\frac{\partial \mL}{\partial (\partial_0 h_{kl})}\frac{\partial (\partial_0 h_{kl})}{\partial(\partial_0 a^{ij})}=
	\nonumber \\
&&=	\frac{\partial \mL}{\partial (\partial_0 h_{kl})}\frac{\partial  h_{kl}}{\partial  a^{ij}}=
	-\frac{\partial \mL}{\partial(\partial_0 h_{kl})}
	h_{km}\frac{\delta h^{mn}}{\delta a^{ij}}h_{nl}=\nonumber \\
&&=
-\frac{1}{\kappa}\sqrt{h}\mG^{klpr}h_{ki}h_{lj}K_{pr}(M^2a)^{\frac{\alpha}{\alpha(D+1)-1}}
-\frac{p_MM}{2}a_{ij} \ .  \nonumber \\
\end{eqnarray}	
Taking the trace of this equation we get
\begin{eqnarray}
&&\pi_{ij}a^{ij}=
\frac{(\alpha -1)}{2\alpha}p_MM 
\nonumber \\
\end{eqnarray}
that implies primary constraint of the theory
\begin{equation}
\mG\equiv \pi_{ij}a^{ij}
-\frac{(\alpha -1)}{2\alpha}p_MM  
\approx 0 \ . 
\end{equation}
Now we are ready to proceed to the definition of the canonical Hamiltonian 
\begin{eqnarray}
&&\mH=\pi_{ij}\partial_0 a^{ij}+\pi_M\partial_0 M-\mL=\frac{\partial \mL}{\partial(\partial_0 h_{ij})}\partial_0 h_{ij}-\mL=\nonumber \\
&&=\frac{N}{\kappa}\sqrt{h}K_{ij}\mG^{ijkl}K_{kl}-\frac{1}{\kappa}N\sqrt{h}r
+\frac{2}{\kappa}\sqrt{h}\mG^{ijkl}K_{kl}\nabla_i N_j \equiv  \nonumber \\
&&\equiv \mH_T+\frac{2}{\kappa}\sqrt{h}\mG^{ijkl}K_{kl}\nabla_i N_j  \ , 
\nonumber \\
\end{eqnarray}
where we used the fact that 
\begin{eqnarray}
&&\pi_{ij}\partial_0 a^{ij}+\pi_M\partial_0 M=\nonumber \\
&&=\frac{\partial \mL}{\partial (\partial_0 h_{kl})}\frac{\partial h_{kl}(a,M)}{\partial a^{ij}}\partial_0 a^{ij}+\frac{\partial \mL}{\partial (\partial_0 h_{kl})}
\frac{\partial h_{kl}(a,M)}{\partial M}\partial_0 M=\nonumber \\
&&=\frac{\partial \mL}{\partial (\partial_0 h_{ij})}\partial_0 h_{ij} \ .  \nonumber \\
\end{eqnarray}
Then using previous relations we obtain explicit form of Hamiltonian in the form 
\begin{eqnarray}
&&\mH_T
=\kappa M^{\frac{2\alpha D-1}{\alpha(D+1)-1}}a^{\frac{1-2\alpha}{2(\alpha(D+1)-1)}}
[\Pi_{mn}a^{mp}a^{nr}\Pi_{pr}
-\frac{1}{D-2}\Pi_{mn}a^{mn}\Pi_{pr}a^{pr}]-\nonumber \\
&&-\frac{1}{\kappa}
(M^2a)^{-\frac{1}{2[\alpha(D+1)-1]}}r(h) \nonumber \\
\end{eqnarray}
where 
\begin{equation}
\Pi_{ij}=\pi_{ij}+\frac{p_MM}{2}a_{ij} \ . 
\end{equation}
Finally we consider last part of the Hamiltonian and after some calculations we get
\begin{eqnarray}
\int d^D\bx \frac{2}{\kappa}\sqrt{h}\mG^{ijkl}K_{kl}\nabla_i N_j=
 \int d^D\bx N^m\mH_m \ , \nonumber \\
\end{eqnarray}
 where $\mH_m$ is defined as
\begin{eqnarray}
\mH_m=2\partial_i[a^{ip}\pi_{pm}]+\partial_m[p_MM]-
2\Gamma^k_{im}a^{ip}\pi_{pk}-\Gamma^k_{km}p_MM \ . 
\end{eqnarray}
%
%
%
In summary we get Hamiltonian $\mH=\mH_T+M^i\mH_i$ 
and set of primary constraints $\pi_i
\approx 0,\mG\approx 0$. In the next section we will study stability of these constraints.
\section{Stability of Primary Constraints}\label{third}
In this section we study stability of the primary constraints. 
We start with the constraints $\pi_m\approx 0$ where the requirement of their preservations implies secondary constraints 
\begin{equation}\label{mHmcons}
\mH_m(\bx)\approx 0 \ . 
\end{equation}
In case of the constraint $\mG\approx 0$ the situation is more involved. 
Recall that  $\mG\approx 0$ has explicit form
\begin{equation}
\mG\equiv \pi_{ij}a^{ij}
-\frac{(\alpha -1)}{2\alpha}p_MM  
\approx 0 \ . 
\end{equation}
This constraint has following Poisson brackets with canonical variables
\begin{eqnarray}
&&\pb{\mG(\bx),a^{ij}(\by)}=-a^{ij}(\bx)\delta(\bx-\by) \ , \quad 
\pb{\mG(\bx),\pi_{ij}(\by)}=\pi_{ij}(\bx)\delta(\bx-\by) \ , \nonumber \\
&&\pb{\mG(\bx),M(\by)}=\frac{\alpha-1}{2\alpha}M\delta(\bx-\by)  \ , \quad 
\pb{\mG(\bx),p_M(\by)}=-\frac{\alpha-1}{2\alpha}p_M(\bx)\delta(\bx-\by) \ 
\nonumber \\
\end{eqnarray}
that implies 
\begin{eqnarray}
&&\pb{\mG(\bx),M^2a(\by)}=
\frac{\alpha(D+1)-1}{\alpha}M^2a(\by)\delta(\bx-\by) \ , 
\quad 
\pb{\mG(\bx),h^{ij}(\by)}=0 \ , \nonumber \\
&&\pb{\mG(\bx),\Pi_{ij}(\by)}=
\Pi_{ij}(\bx)\delta(\bx-\by) \ . \nonumber \\
\end{eqnarray}
We further have
\begin{eqnarray}
&&\pb{\mG(\bx),M^{\frac{2\alpha D-1}{\alpha (D+1)-1}}a^{\frac{1-2\alpha}{2(\alpha(D+1)-1)}}(\by)}=
\nonumber \\
&&=-\frac{1}{2\alpha}M^{\frac{2\alpha D-1}{\alpha (D+1)-1}}a^{\frac{1-2\alpha}{2(\alpha(D+1)-1)}}(\by)\delta(\bx-\by) \ , 
\nonumber \\
&&\pb{\mG(\bx),(M^2a)^{-\frac{1}{2(\alpha(D+1)-1)}}(\by)}=\nonumber \\
&&=-\frac{1}{2\alpha}(M^2a)^{-\frac{1}{2(\alpha(D+1)-1)}}(\by)\delta(\bx-\by) \ .  \nonumber \\
\end{eqnarray}
Collecting these terms together we finally obtain 
\begin{equation}
\pb{\mG(\bx),\mH_T(\by)}=-\frac{1}{2\alpha}\mH_T(\by)\delta(\bx-\by) \ .
\end{equation}
Further, since $\pb{\mG(\bx),\mH_m(\by)}=0$
we immediately find that
\begin{equation}
\partial_t \mG=\pb{\mG,H}=-\frac{1}{2\alpha}\mH_T 
\end{equation}
so that requirement of the preservation of the primary constraint $\mG\approx 0$ implies
secondary constraint 
\begin{equation}
\mH_T\approx 0  \ . 
\end{equation}
At this stage we identified $D+1$ secondary constraints $\mH_i\approx 0 \ , \mH_T\approx 0$ together with $D+1$ primary constraints $\pi_i\approx 0 \ , \mG\approx 0$. 
Now we should check stability of  secondary constraints. To proceed to these calculations it is convenient to express $\mH_i$ in different way. 

We start with following Poisson bracket 
\begin{equation}
\pb{\Pi_{ij}(\bx),h^{kl}(\by)}=-\frac{1}{2}(\delta_i^k\delta_j^l+\delta_i^l\delta_j^k)
\Sigma \delta(\bx-\by) \ , \Sigma=(M^2a)^{\frac{\alpha}{\alpha(D+1)-1}} \ \end{equation}
which has almost canonical form. In order to have Poisson brackets in the canonical form 
we introduce  $\tPi^{ij}$ defined as
\begin{equation}\label{tPi}
\tPi^{ij}=\Sigma^{-1}h^{ik}\Pi_{kl}h^{lj} \ 
\end{equation}
that has following Poisson brackets
\begin{equation}\label{tPihkl}
\pb{\tPi^{ij}(\bx),h_{kl}(\by)}=\frac{1}{2}
(\delta^i_k\delta^j_l+\delta^i_l\delta^j_k)\delta(\bx-\by) \ . 
\end{equation}
Note that $\tPi^{ij}$ is related to $K_{ij}$ by following formula 
\begin{equation}
\tPi^{ij}=-\frac{1}{\kappa}\sqrt{h}\mG^{ijkl}K_{kl} \ .
\end{equation}
Then it is easy to see that the spatial diffeomorphism constraint can be written as 
\begin{equation}
\mH_i=2h_{il}\nabla_k\tPi^{lk} \ . 
\end{equation}
Clearly this diffeomorphism constraint has the same form as spatial diffeomorphism constraint derived  in General Relativity. 
Further, since the Poisson bracket between $\tPi^{ij}$ and $h_{kl}$ (\ref{tPihkl}) has the same form as in case of General Relativity(up to sign) we find that the Poisson brackets between two smeared forms of diffeomorphism constrains has the form
\begin{eqnarray}
\pb{\bT_S(X^i),\bT_S(Y^j)}=\bT_S(X^j\partial_j Y^i-Y^j\partial_j X^i) \ . 
\end{eqnarray}
Let us now turn our attention to the  Hamiltonian constraint $\mH_T$. In fact, using $\tPi^{ij}$ we can write Hamiltonian constraint in the form 
\begin{equation}
\mH_T=\frac{N\kappa}{\sqrt{h}}\tPi^{ij}\mG_{ijkl}\tPi^{kl}-\frac{1}{\kappa}N\sqrt{h}r(h) \ ,
\end{equation}
where $N$ and $h$ are composed from canonical variables $a^{ij}$ and $M$. 
Then it is clear that the Poisson brackets between smeared form of diffeomorphism constraints and Hamiltonian constraint has the same form as in general relativity. 
Explicitly, we have
\begin{eqnarray}
&&\pb{\bT_T(X),\bT_T(Y)}=\bT_S((NX\partial_i(NY)-NY\partial_i(XN))^{ij}) \ ,
\nonumber \\
&&\pb{\bT_S(X^i),\bT_T(X)}=\bT_T(X^i\partial_i(NX)N^{-1}) \ . \nonumber \\
\end{eqnarray}
This result makes an analysis of General Relativity with $a^{ij}$ and $\pi_{ij}$ as canonical variables complete. We see that introducing these new variables 
 leads only to an emergence of new constraint $\mG\approx 0$ which replaces original constraint $p_N\approx 0$. Then the remaining constraint structure is completely the same. Further, from the form of the Hamiltonian and diffeomorphism constraint it is hardly to see that they would simplify resulting Hamiltonian. For that reason we mean that introducing new variables does not bring new benefits for theory.

\section{New Set of Alternative Variables}\label{fourth}
In this section we derive canonical formulation for gravity when we introduce
spatial metric $a^{ij}$ related to $h^{ij}$ in the form
\begin{equation}
a^{ij}=h^\beta h^{ij} \ , \quad h=\det h_{ij}
\end{equation}
while $N$ and $N^i$ remain the  same. As in the second section 
we derive inverse relation between $h^{ij}$ and $a^{ij}$ in the form
\begin{equation}
h^{ij}=a^{ij}a^{-\frac{\beta}{\beta D-1}} \ 
\end{equation}
so that  momentum conjugate to $a^{ij}$ is equal to
\begin{eqnarray}
&&\pi_{ij}=\frac{\delta \mL}{\delta \partial_0 a^{ij}}
=-\frac{\delta \mL}{\delta \partial_0 h_{kl}}
h_{kp}\frac{\delta  h^{pr}}{\delta  a^{ij}}h_{rl}=
\nonumber \\
&&=-\frac{\sqrt{h}}{\kappa}\mG^{klmn}K_{mn}
h_{kp}h_{pr}[\frac{1}{2}(\delta^p_i\delta^r_j+\delta^p_j\delta^r_i)-\frac{\beta}{\beta D-1}a^{pr}a_{ij}]a^{-\frac{\beta}{\beta D-1}} \ . 
\nonumber \\
\end{eqnarray}
As in previous section we find Hamiltonian in the form 
\begin{eqnarray}
\mH=\pi_{ij}\partial_0 a^{ij}-\mL
=N\sqrt{h}K_{ij}\mG^{ijkl}K_{kl}+\frac{N\sqrt{h}}{\kappa}\mG^{klmn}K_{mn}\nabla_kN_l \ . 
\nonumber \\
\end{eqnarray}
Taking the trace of the relation for $\pi_{ij}$ we obtain 
\begin{eqnarray}
\pi
=\frac{\sqrt{h}}{\kappa}\frac{1-D}{\beta D-1}K \ 
\nonumber \\
\end{eqnarray}
and finally we get
\begin{eqnarray}
(\pi_{ij}-\beta\pi a_{ij})h^{im}h^{jn}=-\frac{\sqrt{h}}{\kappa}\mG^{mnkl}K_{kl} a^{-\frac{\beta}{\beta D-1}} \ .
\nonumber \\
\end{eqnarray}
Then the Hamiltonian constrain has the form 
\begin{eqnarray}
\mH_T=\frac{\kappa}{\sqrt{h}}
\Pi^{mn}\mG_{mnkl}\Pi^{kl}-\frac{1}{\kappa}\sqrt{h}r(h) \ , 
\nonumber \\
\end{eqnarray}
where
\begin{equation}
\Pi^{ij}=h^{ik}(\pi_{kl}-\beta\pi a_{kl})h^{lj}a^{\frac{\beta}{\beta D-1}} \ . 
\end{equation}
Then using the fact that
\begin{equation}
\pb{\pi_{ij}-\beta \pi a_{ij},h^{kl}}=
-\frac{1}{2}(\delta^k_i\delta^l_j+\delta^k_j\delta^l_i)
a^{-\frac{\beta}{\beta D-1}} \ 
\end{equation}
it is easy  to see that the Poisson bracket between  $\Pi^{ij}$ and $h_{ij}=a_{ij}
a^{\frac{\beta}{\beta D-1}}$ has the canonical form (up to sign)
\begin{equation}\label{PBPiij}
\pb{\Pi^{ij}(\bx),h_{kl}(\by)}=\frac{1}{2}(\delta^i_k\delta^j_l+\delta^i_l\delta^j_k)
\delta(\bx-\by) \ . 
\end{equation}
Note that the Hamiltonian can be written in the form 
\begin{equation}
\mH=N\mH_T-2\tPi^{ij}\nabla_j N_i\equiv N\mH_T+N^i\mH_i \ ,  \nonumber \\
\end{equation}
where
\begin{equation}
\mH_i=h_{ik}\nabla_j\tPi^{jk} \ . 
\end{equation}
Now as in the case of General Relativity $\pi_N,\pi_i$ which are momenta conjugate to $N$ and $N^i$ are primary constraints of the theory. Then the requirement of their preservation implies that  $\mH_T\approx 0 \ , \mH_i\approx 0$ are secondary constraints that 
have the same form as in the case of General Relativity and also thanks to the Poisson brackets 
(\ref{PBPiij}) we get that the Poisson brackets between these constraints are the same as in General Relativity. In other words $\mH_T\approx 0,\mH_i\approx 0$ are first class constraints. 

Now we show that this theory can be derived from the theory studied in section (\ref{second}) when we fix the gauge symmetry $\mG=\pi_{ij}a^{ij}-\frac{\alpha-1}{2\alpha}p_MM\approx 0$. Let us fix this gauge symmetry by introducing gauge fixing function 
$\mF\equiv M-K\approx 0 \ , K=\mathrm{const}$. Then the Poisson bracket between 
$\mG$ and $\mF$ is non-zero and they are second class constraints that can be explicitly solved. Solving $\mG$ for $p_M$ and $M$ we get
\begin{equation}
p_MM=\frac{2\alpha\pi}{\alpha-1} \ . 
\end{equation}
Inserting this result into $\tPi^{ij}$ defined in (\ref{tPi}) we get 
  that we should identify $\beta$ with $\alpha$ as 
\begin{equation}
\frac{\alpha}{\alpha-1}=-\beta \ . 
\end{equation}
For  $K=1$ we find that  
$h^{ij}=a^{ij}a^{-\frac{\alpha}{\alpha(D+1)-1}}=a^{ij}a^{-\frac{\beta}{\beta D-1}}$ and the correspondence is exact. 

As the final point of this section we express Hamiltonian constraint in terms of physical variables $a^{ij}$ and $\pi_{ij}$
\begin{eqnarray}\label{Hama}
&&\mH_T=\kappa a^{\frac{1}{2(\beta D-1)}}
\left(\pi_{mn}a^{mk}a^{nl}\pi_{kl} +\frac{1}{D-2}(\pi_{ij}a^{ij})^2(-1+4\beta-2\beta^2D)\right)-
\nonumber \\
&&-\frac{1}{\kappa} a^{-\frac{1}{2(\beta D-1)}}r(h) \ . 
	\nonumber \\
\end{eqnarray}	
We see that generally this Hamiltonian constraint  has similar form as in case of the original variables.
On the other hand  we hardly see any simplification introducing new variables $f^{mn}$ defined above. 

{\bf Acknowledgement:}

This work  is supported by the grant “Dualitites and higher order derivatives” (GA23-06498S) from the Czech Science Foundation (GACR).


\end{document}